\documentclass[12pt]{article}
\usepackage{amsfonts}
\usepackage[utf8]{inputenc}
\usepackage{bm}
\usepackage{amsmath}
\usepackage{amssymb}
\usepackage{epsfig}
\usepackage{enumitem}
\usepackage{cite}
\usepackage[bookmarks=true,colorlinks=true,linkcolor=black,citecolor=orange,urlcolor=orange,bookmarksnumbered]{hyperref}
\usepackage{cancel}

\usepackage[dvipsnames]{xcolor}

\setlist[enumerate]{%
wide =0.5\parindent,
listparindent=0pt%
}

\newcommand{\bmat}{\left(\begin{array}}
\newcommand{\emat}{\end{array}\right)}

\def\gtrsim{\mathrel{\raise.3ex\hbox{$>$\kern-.75em\lower1ex\hbox{$\sim$}}}}

\def\-{\hphantom{-}}

\def\s2{\frac{1}{\sqrt2}}

\def\mg{m_{3/2}}
\def\mg2{m^2_{3/2}}

\def\Dsl{\,\raise.15ex\hbox{/}\mkern-13.5mu D} 

\def\be{\begin{equation}}
\def\ee{\end{equation}}
\def\bea{\begin{eqnarray}}
\def\eea{\end{eqnarray}}

\newcommand{\nn}{\nonumber}

\begin{document}


\pagestyle{plain}

\makeatletter
\@addtoreset{equation}{section}
\makeatother
\renewcommand{\theequation}{\thesection.\arabic{equation}}
\pagestyle{empty}
\begin{center}
\ \

\vskip .5cm

\LARGE{\bf Gravitational four-derivative corrections in non-relativistic heterotic supergravity and the $SO(8)$ Green–Schwarz mechanism} \\[10mm]

\vskip 0.3cm

\large{Eric Lescano
 \\[6mm]}

{\small Jozef Stefan Institute, \\ Jamova 39, 1000 Ljubljana, Slovenia\\ [.01 cm]}

{\small \verb"eric.lescano@ijs.si"}\\[1cm]

\small{\bf Abstract} \\[0.5cm]

 \end{center}

We present the first explicit construction of the four-derivative gravitational corrections to heterotic supergravity in the non-relativistic (NR) limit. By extending the Bergshoeff-de Roo identification (BdRi) to NR backgrounds, we obtain the complete finite four-derivative completion of the bosonic gravitational sector of NR heterotic supergravity. We investigate the BdRi in the two currently known non-relativistic limits of heterotic supergravity. In the first case, corresponding to the Lescano-Osten construction, the BdRi gives rise to a gravitational SO(8) Green-Schwarz (GS) mechanism originating from the NR gauge transformation of the Kalb-Ramond field. This mechanism can be trivialized using field redefinitions, and this modification induces a GS mechanism for boost transformations. In the second case, corresponding to the Bergshoeff-Romano construction, the resulting gravitational $SO(8)$ GS mechanism cannot be trivialized, closely paralleling the relativistic heterotic theory. Our results establish a systematic framework for incorporating higher-curvature gravitational dynamics into non-relativistic heterotic backgrounds and provide a first step toward the complete higher-derivative NR heterotic theory.

\setcounter{page}{1}
\pagestyle{plain}
\renewcommand{\thefootnote}{\arabic{footnote}}
\setcounter{footnote}{0}
\newpage
\tableofcontents 
\section{Introduction}

The low-energy effective actions of string theory offer a rich playground for exploring the interplay between geometry, symmetry, and quantum consistency. Among the various string models, heterotic string theory stands out for its unique coupling of gravitational and gauge sectors, where anomaly cancellation via the Green-Schwarz (GS) mechanism \cite{GS} introduces non-trivial modifications to the supergravity limit. At leading order, the heterotic effective action exhibits a close relationship between gravitational and gauge dynamics, a structure that becomes particularly pronounced upon incorporating higher-derivative $\alpha'$-corrections.

Historically, these corrections have been explored via two main approaches. One is the amplitude-based method, whose Lagrangian was constructed by Metsaev-Tseytlin \cite{MT}, in agreement with the string scattering amplitudes \cite{GrossSloan}-\cite{CN}. The other approach, pioneered by Bergshoeff and de Roo, uses the restrictions imposed by ${\cal N}=1$ supersymmetric invariance to construct higher-derivative completions of heterotic supergravity \cite{BdR}. In particular, for the bosonic sector of the theory, Bergshoeff and de Roo found an identification which reveals a profound structural analogy between the gauge and the gravitational sectors, offering a unified treatment of the $\alpha'$-corrections of the theory. This identification has not only clarified the structure of the four-derivative gravitational corrections but has also provided a systematic way to encode the GS anomaly cancellation mechanism in terms of a torsionful spin connection.

In the context of the low-energy limit of non-relativistic string theory \cite{NR1}-\cite{NR8}, the Bergshoeff–de Roo identification (BdRi) has not yet been explored. This extension to the NR regime  provides a systematic framework for organizing $\alpha'$-corrections in a manner consistent with NR symmetries. Since the leading-order NS-NS sector of string theory remains finite under a well-defined NR expansion \cite{NSNS} (due to precise cancellations between divergent terms in the Lagrangian),
\bea
    \hat{g}_{\mu \nu} & = & c^2 {\tau_\mu}^a {\tau_\nu}^b \eta_{ab} + h_{\mu \nu} \, , \label{metricexpansionintro} \\
    \hat{g}^{\mu \nu} & = & \frac{1}{c^2} {\tau^\mu}_a {\tau^\nu}_b \eta^{ab} + h^{\mu \nu} \, , \label{inversemetricexpansionintro} \\
    \hat{B}_{\mu \nu} & = & - c^2 {\tau_\mu}^a {\tau_\nu}^{b}  \epsilon_{ab} + b_{\mu \nu} \label{Bexpansionintro}\, ,  \\
\hat{\phi} & = & \ln(c) + \varphi \, ,
\eea
where $\mu,\nu=1,\dots,10$ are spacetime indices and we have split the flat index $\hat a=(a,a')$ where $a,b, ...=0,1$ are the longitudinal directions and $a^\prime,b^\prime,...=2,\dots,9$ are the transverse ones. The quantities in the previous expansion obey the following Newton-Cartan relations,
\begin{align}
  \tau_{\mu}{}^{a} e^{\mu}{}_{a'} & = \tau^{\mu}{}_{a} e_{\mu}{}^{a'} = 0 \, , \qquad  e_{\mu}{}^{a'} e^{\mu}{}_{b'} = \delta^{a'}_{b'} \, , \\
  \tau_{\mu}{}^{a} \tau^{\mu b} & = \eta^{a b} \, , \qquad \tau_{\mu}{}^{a} \tau^{\nu}{}_{a} + e_{\mu}{}^{a'} e^{\nu}{}_{a'} = \delta_{\mu}^{\nu} \, .
\label{NCrelations}
\end{align}

It is natural to wonder about extensions of this construction to the four-derivative corrections. In this work we focus on the gravitational sector of these corrections, and construct the four-derivative Lagrangian by considering a suitable generalization of the BdRi. We impose this map on the two-derivative terms in the gauge sector (Chern-Simons plus Yang-Mills) of the theory \cite{LescanoHet}. We also impose the BdRi on the pioneering heterotic formulation constructed in \cite{BR}. Since in this formulation $\tau_{\mu}{}^{a}$ transforms under gauge transformations, we limit the identification to the particular case where this transformation is trivial. 

The strategy is to construct the four-derivative gravitational terms from the two-derivative gauge theory, turning gauge contributions into gravitational contributions. For a pedagogical introduction to this technique in the relativistic regime, see \cite{Review}. The main goal of this paper is to construct the NR analogue of the BdRi and use it to derive the four-derivative corrections to the heterotic supergravity action in the NR limit. To this end, we begin by reviewing the relativistic Bergshoeff-de Roo mechanism (section 2), highlighting its reinterpretation of the gravitational $\alpha'$-corrections as arising from a map between non-Abelian gauge fields and torsionful Lorentz connections. We then apply a systematic NR expansion to the bosonic part of the two-derivative heterotic action, including both the NS-NS and gauge sectors, 
 \bea
 \hat{A}_{\mu}^{i} &= c \ {\tau_\mu}^{-} \alpha_{-}^i + \frac{1}{c} a_\mu^i \, ,
\eea
where $v^{\pm}=\frac{1}{\sqrt{2}}(v^{0} \pm v^{1})$, and we construct the NR version of the BdRi in section 3. We first apply the identification in the Lescano-Osten approach: The key technical tool is the light-cone decomposition of the Newton-Cartan geometry, which allows us to isolate the degrees of freedom responsible for the non-trivial transformations of the Kalb-Ramond two-form under local Lorentz transformations. By implementing a field redefinition that isolates the longitudinal component $\tau^-_\mu$ in the GS mechanism, we match the structure of the non-Abelian gauge transformation to the form of the gravitational anomaly cancellation. This redefinition is inspired by the recent results in \cite{Trivial}. Since the $SO(8)$ transformation is ambiguous in this limit, instead of eliminating the whole transformation one can remove only the longitudinal component, which is the key step implemented to perform the BdRi. This leads us to identify the NR gauge field with an $SO(8)$ spin connection with torsion, and to express the Chern-Simons-like contributions to the three-form field strength in terms of the gravitational degrees of freedom. The finite action is systematically computed following the previous steps. 

We then apply the identification to the Bergshoeff-Romano approach in section 3.2. For doing so, we work in the limit where the vielbeins do not receive contributions in their gauge transformations. In this scenario, the gauge GS mechanism cannot be trivialized by field redefinitions and therefore the BdRi produces a gravitational GS mechanism which cannot be trivialized, similar to what happens in the relativistic case. The resulting four-derivative actions contain finite terms depending on the curvature $R^{(-)}_{\mu\nu a'b'}$ and on the intrinsic torsion $\tau_{\mu \nu}{}^{+}$. By construction the full actions are invariant under $SO(8)$, boost transformations and diffeomorphism invariance, all of them preserved by the BdRi. In particular, the new four-derivative terms are not $SO(8)$ invariant, but they compensate the transformation of the two-derivative action with respect to the $SO(8)$ GS mechanism. In both setups, the boost transformations are a symmetry of the theory, but higher-derivative corrections for the boost transformations need to be included. We explicitly compute the boost transformations in presence of gauge symmetry in section 4, and its gBdRi in section 5. 

In section 6 we discuss the possibility of constructing the Metsaev and Tseytlin action \cite{MT} for the NR heterotic supergravity with four-derivative terms. This formulation it is promising because it could provide the form of the four-derivative bosonic corrections, which are not guided by a GS mechanism.  Finally, in section 7, we conclude the work and discuss some potential long-term lines of research.

\section{Review: The Bergshoeff-de Roo identification}

The gravitational heterotic action principle, up to four-derivative terms, can be written as \cite{BdR}
\bea
S_{\rm het} = \int d^Dx \sqrt{- \hat g} e^{-2 \hat  \phi} (\hat R + 4 \partial_\mu \hat  \phi \partial^{\mu} \hat  \phi - \frac{1}{12} \tilde H_{\mu \nu \lambda} \tilde H^{\mu \nu \lambda} + \frac18 \hat  R^{(-)}_{\mu \nu}{}^{ab} \hat  R^{(-)\mu \nu}{}_{ab})
\label{LagHetL}
\eea
where $\tilde H_{\mu \nu \rho}= 3\left(\partial_{[\mu}\hat B_{\nu\rho]}+\frac12 \hat C^{(-)}_{\mu\nu\rho}\right) = \hat  H_{\mu \nu \rho} +\frac32 \hat  C^{(-)}_{\mu\nu\rho} \,$ and the Chern-Simons 3-form is given by
\bea
\label{CSdef}
\hat  C^{(-)}_{\mu \nu \rho} & = & \hat  w^{(-)}_{[\mu}{}^{ab} \partial_{\nu} \hat  w^{(-)}_{\rho]ab} + \frac23 \hat  w^{(-)}_{[\mu}{}^{ab} \hat  w^{(-)}_{\nu b}{}^{c} \hat  w^{(-)}_{\rho] ca} \, .
\eea
In the previous expression we used a torsionful spin-connection,
\bea
\hat  w^{(-)}_{\mu a b} = \hat  w_{\mu a b} - \frac12 \hat  H_{\mu \nu \rho} e^{\nu}{}_{a} e^{\rho}{}_{b} \, ,
\eea
so, for example, the Riemann tensor in the action is given by,
\bea
\hat  R^{(-)}_{\mu \nu a b} = - 2 \partial_{[\mu} \hat  w^{(-)}_{\nu] a b} + 2 \hat  w^{(-)}_{[\mu a}{}^{c} \hat  w^{(-)}_{\nu] c b} \, .
\eea

The invariance of the action with respect to Lorentz transformations requires a gravitational GS mechanism of the form
\bea
\delta_{\Lambda} \hat  B_{\mu \nu} = - \frac12 \partial_{[\mu} \Lambda^{ab} \hat  w^{(-)}_{\nu]ab} \, . 
\label{Lorentz1}
\eea
This transformation cannot be trivialized by considering field redefinitions on $B_{\mu \nu}$, in the relativistic case.

The key observation of Bergshoeff and de Roo \cite{BdR} is that the structure of the gravitational four-derivative action resembles the form of the two-derivative Yang-Mills plus Chern-Simons theory coupled to the NS-NS sector of string theory, 
\bea
S_{\rm{het}} = \int d^{D}x \sqrt{- \hat  g} e^{-2 \hat  \phi} (\hat  R + 4 \partial_{\mu} \hat  \phi \partial^{\mu} \hat  \phi - \frac{1}{12} \bar{H}_{\mu \nu \rho} \bar{H}^{\mu \nu \rho} - \frac14 F_{\mu \nu}{}^{i} F^{\mu \nu}{}_{i}) \, ,
\label{het0}
\eea
where 
\be
\bar H_{\mu\nu\rho}=3\left(\partial_{[\mu}\hat  B_{\nu\rho]}- \hat  C_{\mu\nu\rho}^{(g)}\right)\, , \label{barH}
\ee 
and $\hat  C_{\mu\nu\rho}^{(g)}$ is the gauged Chern-Simons 3-form, defined as
\be
\hat  C_{\mu\nu\rho}^{(g)}=\hat  A^i_{[\mu}\partial_\nu \hat  A_{\rho]i}-\frac13 f_{ijk}A_\mu^i \hat  A_\nu^j \hat  A_\rho^k \, . 
\ee

For this reason, the authors designed a map to identify the gauge degrees of freedom with gravitational degrees of freedom (see \cite{Review} for a comparison with other actions, such as the Metsaev and Tseytlin action \cite{MT}). 

Let us define the gauge generators $(t_{i})^{ab}$ such that
\bea
(t_{i})^{ab} (t^{i})_{cd} & = & - \delta_{[c}^{a} \delta_{d]}^{b} \, , \\
(t^{i})_{ab} (t_{j})^{ab} & = & - \delta_{j}^i \, , \\
f_{i}{}^{kl} (t^i)_{ab} & = & 2 \sqrt{2} (t^k)_{[a|c} (t^l)^{c}{}_{b]} \, .
\label{map}
\eea
The gauge generators are antisymmetric in their flat indices, $(t_{i})^{ab}=-(t_{i})^{ba}$ and each generic gauge vector/parameter can be written in the following way,
\bea
v^{i} & = & - v^{ab} (t^i)_{ab} \, , \\
\lambda^i & = & - \lambda^{ab} (t^i)_{ab} \, .
\eea
If we apply the previous relations to the non-Abelian gauge transformation of the B-field, $\delta_{\lambda} \hat B_{\mu\nu} =  -(\partial_{[\mu} \lambda^i) \hat  A_{\nu]i}$,  we find
\bea
\delta_{\lambda} \hat  B_{\mu \nu} = \partial_{[\mu} \lambda_{a b} \hat  A_{\nu]}{}^{a b} \, .
\eea
We now identify 
\bea
\label{sugraidl}
\lambda_{ab} & = & - \frac{1}{\sqrt 2}\Lambda_{ab} \, , \\ \hat  A_{\nu}{}^{a b} & = & \frac{1}{\sqrt 2} \hat  w^{(-)}_{\nu}{}^{ab} \, ,
\label{sugraidA}
\eea
obtaining the Lorentz GS transformation from the gauge GS transformation $\delta_{\lambda} B_{\mu \nu} \rightarrow \delta_{\Lambda} B_{\mu \nu}$.
Moreover if we apply (\ref{map}) at the level of the action considering the identification (\ref{sugraidA}), it is straightforward to find
\bea
\label{Rrel}
\hat  F_{\mu \nu}{}^{ab} & 
\rightarrow & - \frac{1}{\sqrt 2} \hat  R^{(-)}_{\mu \nu}{}^{ab} \, , \\ \label{Crel}
3(\partial_{[\mu} \hat  B_{\nu \rho]} - \hat  C^{(g)}_{\mu \nu \rho}) & \rightarrow & 3 (\partial_{[\mu} \hat  B_{\nu \rho]} + \frac12 \hat  C^{(-)}_{\mu \nu \rho}) \, .
\eea
In the next section we will use this procedure, but after taking the NR limit in the heterotic setup.

\section{The non-relativistic version of the Bergshoeff-de Roo identification}
We consider a NR expansion of the form \cite{NSNS},
\bea
    \hat{g}_{\mu \nu} & = & c^2 {\tau_\mu}^a {\tau_\nu}^b \eta_{ab} + h_{\mu \nu} \, , \label{metricexpansion} \\
    \hat{g}^{\mu \nu} & = & \frac{1}{c^2} {\tau^\mu}_a {\tau^\nu}_b \eta^{ab} + h^{\mu \nu} \, , \label{inversemetricexpansion} \\
    \hat{B}_{\mu \nu} & = & - c^2 {\tau_\mu}^a {\tau_\nu}^{b}  \epsilon_{ab} + b_{\mu \nu} \label{Bexpansion}\, ,  \\
\hat{\phi} & = & \ln(c) + \varphi \, ,
\eea 
While a priori this expansion produces divergent tensorial quantities upon taking the NR limit $c \rightarrow \infty$ (i.e. Riemann tensor and its traces, the H-field strength, etc.), the NS-NS supergravity Lagrangian,
\bea
S = \int d^{10}x \sqrt{-\hat g} e^{-2 \hat \phi} \Big( \hat R + 4 \partial_{\mu} \hat \phi \partial^{\mu} \hat \phi - \frac{1}{12} \hat H_{\mu \nu \rho} \hat H^{\mu \nu \rho}\Big)  \, ,
\label{Sbos0}
\eea
with 
\bea
\hat H_{\mu \nu \rho} =  3 \partial_{[\mu} \hat B_{\nu \rho]} \, ,
\eea
remains finite due to a curious cancellation between the divergent part of the Ricci scalar and the divergent part of the $\hat H$ squared term,
\bea
L= e e^{-2 \varphi} (R^{(0)} + 4 \partial_{\mu}\varphi  \partial_{\nu} \varphi h^{\mu \nu} + T_{H^2}(H,\tau,h))
\eea
where
\bea
T_{H^2}(H,\tau,h) & = & 2 \eta_{a b} \eta^{c d} \nabla_{[\mu}{\tau_{\nu]}{}^{a}} \nabla_{[\rho}{\tau_{\sigma]}{}^{b}} \tau^{\mu}{}_{c} \tau^{\sigma}{}_{d} h^{\nu \rho} + 2 \epsilon_{a b} \epsilon_{c d} \eta^{a e} \eta^{c f} \nabla_{[\mu}{\tau_{\nu]}{}^{b}} \nabla_{[\rho}{\tau_{\sigma]}{}^{d}} \tau^{\nu}{}_{f} \tau^{\sigma}{}_{e} h^{\mu \rho} \nn \\ && - \frac{1}{12} H_{\mu \nu \rho} H_{\sigma \gamma \epsilon} h^{\mu \sigma} h^{\nu \gamma} h^{\rho \epsilon} + \epsilon_{a b} \eta^{a c} \nabla_{[\mu}{\tau_{\nu]}{}^{b}} H_{\rho \sigma \gamma} \tau^{\sigma}{}_{c} h^{\mu \rho} h^{\nu \gamma} \nn \\
\eea
and in the previous expression we used $H_{\mu \nu \rho}=3\partial_{[\mu}b_{\nu \rho]}$.

\subsection{The gravitational non-relativistic heterotic action up to four-derivatives in the Lescano-Osten approach}

The inclusion of the gauge vector of the heterotic supergravity can be easily done by considering the following expansion \cite{EandD}
 \bea
 \hat{A}_{\mu}^{i} &= c \ {\tau_\mu}^{-} \alpha_{-}^i + \frac{1}{c} a_\mu^i \, ,
\label{Aexpansion}
\eea
where $v^{\pm}=\frac{1}{\sqrt{2}}(v^{0} \pm v^{1})$, and the
full action is
\bea
S_{\rm{het}} = \int d^{10}x \sqrt{-\hat g} e^{-2 \hat \phi} (\hat R + 4 \partial_{\mu} \hat \phi \partial^{\mu} \hat \phi - \frac{1}{12} \hat{H}_{\mu \nu \rho} \hat{H}^{\mu \nu \rho} - \frac14 \hat F_{\mu \nu}{}^{i} \hat F^{\mu \nu}{}_{i}) \, ,
\label{Shet0}
\eea
with
\be
\hat H_{\mu\nu\rho}=3\left(\partial_{[\mu}\hat B_{\nu\rho]}- \hat C_{\mu\nu\rho}^{(g)}\right)\, , 
\ee 
\be
\hat C_{\mu\nu\rho}^{(g)}= \hat A^i_{[\mu}\partial_\nu \hat A_{\rho]i}-\frac13 \hat f_{ijk} \hat A_\mu^i \hat A_\nu^j \hat A_\rho^k \, ,
\ee
\bea
\hat F_{\mu \nu}{}^{i} = 2 \partial_{[\mu} \hat A_{\nu]}{}^{i} - \hat f^{i}{}_{jk} \hat A_{\mu}{}^{j} \hat A_{\nu}{}^{k} \, ,
\eea
with $\hat f_{i j k}= c f_{i j k}$. In this case, the Lagrangian also remains finite after taking the limit $c \rightarrow \infty$ \cite{LescanoHet} and it can be written as 
\bea
L= e e^{-2 \varphi} \Big(R^{(0)} + 4 \partial_{\mu}\varphi  \partial_{\nu} \varphi h^{\mu \nu} + T_
{H^2}(\bar H,\tau,h,\alpha) + T_{F}(\tau,h,\alpha,F)\Big)
\eea
where
\bea
&& T_{H^2}(\bar H,\tau,h,\alpha) =  \frac{1}{2} \alpha_{-}^{2}  \eta^{a b} \nabla_{[\mu}{\tau_{\nu]}^{-}} \bar H_{\rho \sigma \gamma} \tau_{\epsilon}^{-} \tau^{\rho}{}_{a} \tau^{\epsilon}{}_{b} h^{\mu \sigma} h^{\nu \gamma} + 2 \alpha_{-}^{2} \epsilon_{a b} \eta^{a c} \eta^{d e} \nabla_{[\mu}{\tau_{\nu]}^{b}} \nabla_{[\rho}{\tau_{\sigma]}^{-}} \tau_{\gamma}^{-} \tau^{\mu}_{d} \tau^{\sigma}_{c} \tau^{\gamma}_{e} h^{\nu \rho} \nn \\ && - \frac{1}{2} \alpha_{-}^{4} \eta^{a b} \eta^{c d} \nabla_{[\mu}{\tau_{\nu]}^{-}} \nabla_{[\rho}{\tau_{\sigma]}^{-}} \tau_{\gamma}^{-} \tau_{\epsilon}^{-} \tau^{\mu}_{a} \tau^{\sigma}_{c} \tau^{\gamma}_{d} \tau^{\epsilon}_{b} h^{\nu \rho} + 2 \eta_{a b} \eta^{c d} \nabla_{[\mu}{\tau_{\nu]}{}^{a}} \nabla_{[\rho}{\tau_{\sigma]}{}^{b}} \tau^{\mu}{}_{c} \tau^{\sigma}{}_{d} h^{\nu \rho} \nn \\ && + 2 \epsilon_{a b} \epsilon_{c d} \eta^{a e} \eta^{c f} \nabla_{[\mu}{\tau_{\nu]}{}^{b}} \nabla_{[\rho}{\tau_{\sigma]}{}^{d}} \tau^{\nu}{}_{f} \tau^{\sigma}{}_{e} h^{\mu \rho} - \frac{1}{12} \bar H_{\mu \nu \rho} \bar H_{\sigma \gamma \epsilon} h^{\mu \sigma} h^{\nu \gamma} h^{\rho \epsilon} \nn \\ && - \epsilon_{a b} \eta^{a c} \nabla_{\mu}{\tau_{\nu}{}^{b}} \tau^{\rho}{}_{c} \bar H_{\rho \sigma \gamma} h^{\mu \sigma} h^{\nu \gamma} 
\label{Hcontributions}
\eea
is the finite contribution coming from the $\hat H^2$-term, with $\alpha_{-}^2=\alpha_{-}^{i} \alpha_{-}^{j} \eta_{i j}$ and the gauge invariant 3-form is given by 
\bea
\bar H_{\mu \nu \rho} =  H_{\mu \nu \rho} + 3 \partial_{[\mu}(a_{\nu}^{i} \tau_{\rho]}^{-} \alpha_{-i}) \, .
\eea

The finite contribution coming from the $\hat F^2$-term is given by
\bea
T_{F}(\tau,h,\alpha,F) & = & - 2 \alpha_{-}^{2} \eta^{a b} \nabla_{[\mu}{\tau_{\nu]}^{-}} \nabla_{[\rho}{\tau_{\sigma]}^{-}} \tau^{\nu}_{a} \tau^{\sigma}_{b} h^{\mu \rho} - 2 \alpha_{-}^{i} \eta^{a b} \eta_{i j} \nabla_{\mu}{\alpha_{-}^{j}} \nabla_{[\nu}{\tau_{\rho]}^{-}} \tau_{\sigma}^{-} \tau^{\rho}_{a} \tau^{\sigma}_{b} h^{\mu \nu} \nn \\ &&  - F_{\mu \nu}^{i} \alpha_{-}^{j} \eta_{i j} \nabla_{\rho}{\tau_{\sigma}^{-}} h^{\mu \rho} h^{\nu \sigma} - \frac{1}{2} \eta^{a b} \eta_{i j} \nabla_{\mu}{\alpha_{-}^{i}} \nabla_{\nu}{\alpha_{-}^{j}} \tau_{\rho}^{-} \tau_{\sigma}^{-} \tau^{\rho}_{a} \tau^{\sigma}_{b} h^{\mu \nu} , 
\label{Fcontributions}
\eea
where $F_{\mu \nu}{}^{i}= 2 \partial_{[\mu} a_{\nu]}^{i} - f^{i}{}_{j k} a_{\mu}^{j} a_{\nu}^{k}$ and the gauge covariant derivative is defined as
\bea
\nabla_{\mu} v^{i} = \partial_{\mu} v^{i} - f^{i}{}_{jk} a_{\mu}{}^{j} v^{k} \, .
\eea 

The non-Abelian gauge transformations after considering the limit $c\rightarrow \infty$ for the fundamental fields are
\bea
\delta_{\lambda} a_\mu^i & = & \partial_\mu \lambda^i + f^i{}_{jk} \lambda^j a_\mu^k\, , \label{gaugetransa} \\
\delta_{\lambda} \alpha_{-}^i & = & f^i{}_{jk} \lambda^j \alpha_{-}^k\, , \label{gaugetransalpha} \\ 
\delta_{ \lambda} b_{\mu\nu} & = & -(\partial_{[\mu} \lambda^i) \alpha_{-i} \tau_{\nu]}^-\, , \label{gaugetransb} \, .
\eea
   
On the other hand, the gravitational and torsionful $SO(8)$ GS mechanism after taking the limit is given by  \footnote{This expression includes the torsion in the SO(8) GS mechanism and agrees with the result in \cite{Trivial}, which only contains the torsionless part.},
\bea
\delta_{\Lambda} b_{\mu \nu} = - \frac{1}{4} \partial_{[\mu} \Lambda^{a' b'} \tau_{\nu]}{}^{d} (\eta_{d e} + \frac23 \epsilon_{d e}) \tau_{\rho \sigma}{}^{e}  e^{\rho}{}_{a'} e^{\sigma}{}_{b'} \, .
\eea
The previous transformation can be trivialized \cite{Trivial} by considering a suitable field redefinition, at cost of including a new higher-derivative boost transformation on the b-field (explicitly computed in section \ref{Sec:newboosts}). Instead of canceling the entire transformation, we will consider the light-cone  redefinition of one of the contributions
\bea
\bar b_{\mu \nu} = b_{\mu \nu} + \frac{5}{12} S^{(-)}_{[\mu}{}^{a' b'} \tau_{\nu]}{}^{+}  \tau_{\rho \sigma}{}^{-}  e^{\rho}{}_{a'} e^{\sigma}{}_{b'} \, ,
\eea
where the $SO(8)$ connection with torsion is defined as
\bea
S^{(-)}{}_{\mu}{}^{a' b'} = - 2 e^{\nu [a'} \partial_{[\mu} e_{\nu]}{}^{b']} + e_{\mu c'} e^{\nu a'} e^{\rho b'} \partial_{[\nu} e_{\rho]}{}^{c'} - \frac12 \tau_{\mu}{}^{a} \epsilon_{a b} \tau^{\nu b} e^{\rho a'} e^{\sigma b'} H_{\nu \rho \sigma} \, .
\eea

In this case, the $SO(8)$ GS mechanism is given by
\bea
\delta_{\lambda} \bar b_{\mu \nu} = \frac{5}{12} \partial_{[\mu} \Lambda^{a' b'} \tau_{\nu]}{}^{-}  \tau_{\rho \sigma}{}^{+}  e^{\rho}{}_{a'} e^{\sigma}{}_{b'} \, .
\eea
The strategy implemented with the previous field redefinition is to isolate a $\tau_{\nu]}{}^{-}$ in the gravitational mechanism, so that it has the same structure of (\ref{gaugetransb}). 

We now identify 
\bea
\label{NRmap}
\lambda_{a'b'} & = & - \frac{1}{\sqrt 2}\Lambda_{a'b'} \, , \\ \alpha_{- a' b'} & = & - \frac{5 \sqrt2}{12} \tau_{\rho \sigma}{}^{+} e^{\rho}{}_{a'} e^{\sigma}{}_{b'} \equiv \beta_{+ a' b'} \, .
\eea
obtaining the Lorentz GS transformation from the gauge GS transformation. Also, analyzing the NR gauge transformation of $a_{\mu}{}^{i}$, we need to identify this object with a $SO(8)$ connection with torsion. Therefore, the NR BdRi for the gauge connection is
\bea
a_{\mu}{}^{a' b'} = \frac{1}{\sqrt{2}} S^{(-)}{}_{\mu}{}^{a' b'}
\eea
and the curvature of the gauge connection is now identified as
\bea
F_{\mu \nu a' b'} = \sqrt{2} \partial_{[\mu} S^{(-)}_{\nu] a' b'} - {\sqrt 2} S^{(-)}_{[\mu a'}{}^{c'} S^{(-)}_{\nu] c' b'} \equiv -\frac{1}{\sqrt 2} R^{(-)}_{\mu \nu a' b'}\, .
\eea
With all the previous identifications, in the next section we compute the finite four-derivative gravitational corrections to the action (\ref{Sbos0}) in the NR limit.

We impose the BdRi in the gauge contributions of (\ref{Hcontributions}) and (\ref{Fcontributions}). The expression for $\bar H_{\rho \mu \nu}$ after the identifications is given by, 
\bea
\bar H_{\rho \mu \nu} = 3 \partial_{[\rho} (b_{\mu \nu]} + \frac{5}{12} S^{(-)}_{\mu a' b'} \tau_{\rho \sigma}{}^{+} e^{\rho}{}_{a'} e^{\sigma}{}_{b'} \tau_{\nu]}{}^{-}) \, .
\eea
Similarly, we can find the form of $\alpha_{-}^2$ after the identifications,
\bea
\alpha_{-}^2 = - \frac{25}{72} \tau_{\rho \sigma}{}^{+} \tau_{\mu \nu}^{+} h^{\mu \rho} h^{\nu \sigma} \equiv -\beta_{+}^2 \, .
\eea
Since the previous contribution already has two derivatives, it is not necessary to compute $\alpha_{-}^4$, this contribution will correspond to a $\alpha'_{NR}{}^2$ contribution. However, as we can see, the procedure is predicting a higher-order structure above the $\alpha'_{NR}$ order, similarly to what happens in the relativistic case. 

The full four-derivative gravitational contribution is therefore given by
\bea
L= e e^{-2 \varphi} \Big(R^{(0)} + 4 \partial_{\mu}\varphi  \partial_{\nu} \varphi h^{\mu \nu} + T_
{H}(\bar H,\tau,h,\beta) + T_{R}(\tau,h,\beta,R)\Big)
\eea
where the corrections coming from the Chern-Simons structure of the action are given by
\bea
T_{H}(\bar H,\tau,h,\beta) & = &  - \frac{1}{2} \beta_{+}^{2}  \eta^{a b} \nabla_{[\mu}{\tau_{\nu]}^{-}} \bar H_{\rho \sigma \gamma} \tau_{\epsilon}^{-} \tau^{\rho}{}_{a} \tau^{\epsilon}{}_{b} h^{\mu \sigma} h^{\nu \gamma} \nn \\ && - 2 \beta_{+}^{2} \epsilon_{a b} \eta^{a c} \eta^{d e} \nabla_{[\mu}{\tau_{\nu]}^{b}} \nabla_{[\rho}{\tau_{\sigma]}^{-}} \tau_{\gamma}^{-} \tau^{\mu}_{d} \tau^{\sigma}_{c} \tau^{\gamma}_{e} h^{\nu \rho} \nn \\ && + 2 \eta_{a b} \eta^{c d} \nabla_{[\mu}{\tau_{\nu]}{}^{a}} \nabla_{[\rho}{\tau_{\sigma]}{}^{b}} \tau^{\mu}{}_{c} \tau^{\sigma}{}_{d} h^{\nu \rho} \nn \\ && + 2 \epsilon_{a b} \epsilon_{c d} \eta^{a e} \eta^{c f} \nabla_{[\mu}{\tau_{\nu]}{}^{b}} \nabla_{[\rho}{\tau_{\sigma]}{}^{d}} \tau^{\nu}{}_{f} \tau^{\sigma}{}_{e} h^{\mu \rho} 
\nn \\ && 
- \frac{1}{12} \bar H_{\mu \nu \rho} \bar H_{\sigma \gamma \epsilon} h^{\mu \sigma} h^{\nu \gamma} h^{\rho \epsilon} \nn \\ && - \epsilon_{a b} \eta^{a c} \nabla_{\mu}{\tau_{\nu}{}^{b}} \tau^{\rho}{}_{c} \bar H_{\rho \sigma \gamma} h^{\mu \sigma} h^{\nu \gamma}
\label{Hcontributionsnew}
\eea
while the four-derivative corrections associated to the Riemann tensor $R^{(-)}_{\mu \nu a' b'}$ are given by
\bea
T_{R}(\tau,h,\beta,R) & = & 2 \beta_{+}^{2} \eta^{a b} \nabla_{[\mu}{\tau_{\nu]}^{-}} \nabla_{[\rho}{\tau_{\sigma]}^{-}} \tau^{\nu}_{a} \tau^{\sigma}_{b} h^{\mu \rho} + 2 \beta_{+}^{a' b'} \eta^{a b} \nabla_{\mu}{\beta_{+ a' b'}} \nabla_{[\nu}{\tau_{\rho]}^{-}} \tau_{\sigma}^{-} \tau^{\rho}_{a} \tau^{\sigma}_{b} h^{\mu \nu} \nn \\ &&  - \frac{1}{\sqrt{2}} R_{\mu \nu}^{(-)a' b'} \beta_{+ a' b'} \eta_{i j} \nabla_{\rho}{\tau_{\sigma}^{-}} h^{\mu \rho} h^{\nu \sigma} + \frac{1}{2} \eta^{a b} \nabla_{\mu}{\beta_{+}^{a' b'}} \nabla_{\nu}{\beta_{+ a' b'}} \tau_{\rho}^{-} \tau_{\sigma}^{-} \tau^{\rho}_{a} \tau^{\sigma}_{b} h^{\mu \nu} \, , \nn \\
\label{Fcontributionsnew}
\eea
where the covariant derivatives now contain the $SO(8)$ connection, providing covariant quantities under this symmetry, together with the gravitational GS mechanism.  Finally, the transformations of the fundamental fields are
\bea
\delta e^{\mu}{}_{a'} & = & L_{\xi} e^{\mu}{}_{a'}  + \Lambda_{a'}{}^{b'} e^{\mu}{}_{b'}     \, , \\
\delta \tau_{\mu}{}^{a} & = & L_{\xi} \tau_{\mu}{}^{a} + \Lambda^{a}{}_{b} \tau_{\mu}{}^{b}  \, , \\
\delta \varphi & = & L_{\xi} \varphi \, , \\
\delta \bar b_{\mu\nu} & = & L_{\xi} \bar b_{\mu \nu} + 2 \partial_{[\mu} \zeta_{\nu]} - 2 \epsilon_{a b} \lambda^{a}{}_{a'} \tau_{[\mu}{}^{b} e_{\nu]}{}^{a'} \nn \\ &&
+ \frac{5}{12} \partial_{[\mu} \Lambda^{a' b'} \tau_{\nu]}{}^{-}  \tau_{\rho \sigma}{}^{+}  e^{\rho}{}_{a'} e^{\sigma}{}_{b'} + \Delta^{(1)}_{boost} b_{\mu \nu} \, ,
\eea
where $\xi^{\mu}$ is a parameter of infinitesimal diffeomorphisms, $\Lambda_{a b}$ is a $SO(1,1)$ parameter and $\Lambda_{a'b'}$ is a parameter of $SO(8)$ transformations. The action of heterotic supergravity up to four-derivative terms is invariant under all these symmetries plus boost transformations, which we have not explicitly included in the previous list since they would need higher-derivative contributions in the transformations (for example, in the b-field transformation). However, the boost transformations are a symmetry of the NR generalized Bergshoeff-de Roo action since the two-derivative gauge Lagrangian is boost invariant. We will explore the explicit form of the higher-derivative boost corrections in section \ref{Sec:newboosts}.  

We recall that the final contribution to the $\bar b_{\mu \nu}$ transformation can be eliminated by using field-redefinitions, as shown in \cite{Trivial}. Imposing this redefinition would simplify the transformations and make them compatible with the transformations of the bosonic supergravity in the Metsaev and Tseytlin approach \cite{MT} (see section 4). An alternative identification is given in section 3.2, by applying the identification in a particular case of the formulation given in \cite{BR}, paying the price that the GS mechanism cannot be trivialized, as it happens in the relativistic case. At this point it's important to recall that the trivialization of the $SO(8)$ GS mechanism in the Lescano-Osten approach induce higher-derivative boost transformations, which are discussed in section 4 and 5.

\subsection{The gravitational non-relativistic heterotic action up to four-derivatives in the Bergshoeff-Romano approach}

Following the construction \cite{BR}, it is possible to construct an alternative finite heterotic theory in the NR limit by considering,
\bea
\hat V_{\nu}{}^{i} = c^2 \tilde \tau_{\nu}{}^{-} v_{-}{}^{i} + \tilde \tau_{\nu}{}^{+} v_{+}{}^{i} + \tilde e_{\nu}{}^{a'} v_{a'}{}^{i} \, ,
\eea
where $\hat V_{\mu}{}^{i}$ plays the role of the $\hat A_{\mu}{}^{i}$ vector. The c-expansion for the vielbein in this setup is given by  
\bea
\hat E_{\mu}{}^{-} & = & c \tilde \tau_{\mu}{}^{-}, \\ \hat E_{\mu}{}^{+} & = & - c^3 \frac{v_{-}{}^{i} v_{- i}}{2} \tilde \tau_{\mu}{}^{-} + c \tilde \tau_{\mu}{}^{+} \, , \\ \hat E_{\mu}{}^{a'} & = & \tilde e_{\mu}{}^{a'} \, ,
\eea
while the expansion for the B-field is
\bea
\hat B_{\mu \nu} = c^2 \tilde \tau_{\mu}{}^a {\tilde \tau_\nu}^{b}  \epsilon_{ab} + 2 c^2 \tilde \tau_{[\mu}{}^{-} \tilde e_{\nu]}{}^{a'} v_{-}{}^{i} v_{a' i} + \tilde b_{\mu \nu} \, .
\eea
The gauge transformations of the gauge degrees of freedom are given by
\bea
\delta_{\lambda} v_{+}{}^{i} & = & \partial_{+} \lambda^{i} + \sqrt{2} g f_{jk}{}^{i} \lambda^{j} v_{+}{}^{k} \, , \\
\delta_{\lambda} v_{-}{}^{i} & = & \sqrt{2} g f_{jk}{}^{i} \lambda^{j} v_{-}{}^{k} \, , \\
\delta_{\lambda} v_{a'}{}^{i} & = & \partial_{a} \lambda^{i} + \sqrt{2} g f_{jk}{}^{i} \lambda^{j} v_{a'}{}^{k} \, . 
\eea

In this prescription, the vielbein receives gauge contributions in its expansion so $\tilde \tau_{-}{}^{a}$ is gauge invariant but $\tilde \tau_{+}{}^{a}$ transforms in a non-covariant way under gauge transformations,
\bea
\delta \tilde \tau_{\mu}{}^{+} = \frac{v_{-}{}^{i} \partial_{-} \lambda^{i}}{1+v_{+}{}^{i} v_{-i}} \tilde \tau_{\mu}{}^{-} \, .
\eea
Since $v_{-}{}^{i}$ is boost invariant and it transforms as a vector under gauge transformations, we will work with the case $v_{-}{}^{i}=0$, so that 
\bea
\tilde \tau_{\mu}{}^{a} = \tau_{\mu}{}^{a} \quad \quad \tilde e_{\mu}{}^{a'} = e_{\mu}{}^{a'} \, . 
\eea
The gauge GS mechanism for the b-field in this case is given by
\bea
\delta \tilde b_{\mu \nu} & = & 2 \tau_{[\mu}{}^{+} e_{\nu]}{}^{a'} (v_{+}{}^i \partial_{a'}\lambda_i - v_{a' i} \partial_{+}\lambda^{i}) - 2 e_{[\mu}{}^{a'} e_{\nu]}{}^{b'} v_{a'}{}^{i} \partial_{b'} \lambda_{i} \nn \\ && - 4 (\tau_{[\mu}{}^{-} \tau_{\nu]}{}^{+} v_{+i} + \tau_{[\mu}{}^{-} e_{\nu]}{}^{a'} v_{a'i} )\partial_{-} \lambda^{i} \, .
\label{GSBR}
\eea
The previous transformation is not ambiguous and, therefore, it cannot be eliminated with a field redefinition. When $v_{-}{}^{i}\neq 0$, one can try to use the anomalous transformation of $\tilde \tau_{\mu}{}^{+}$ to trivialize the GS mechanism. In the case $v_{-}{}^{i}=0$ the identification for the connections $v_{ai}$ and $v_{+i}$ follows the same logic as in Section 3. These fields are identified with suitable rotations of the $SO(8)$ connection, 
\bea
v_{a' b' c'} = \frac{1}{\sqrt{2}} e^{\mu}{}_{a'} S^{(-)}_{\mu b' c'} \equiv \frac{1}{\sqrt{2}} S^{(-)}_{a' b' c'}
\eea
and
\bea
v_{+ b' c'} = \frac{1}{\sqrt{2}} \tau^{\mu}{}_{+} S^{(-)}_{\mu b' c'} \equiv \frac{1}{\sqrt{2}} S^{(-)}_{+ b' c'} \, .
\eea
The previous identifications can be applied to the mechanism (\ref{GSBR}) to obtain the gravitational GS mechanism, compatible with this formulation. The result  is given by
\bea
\delta \tilde b_{\mu \nu} & = &  \tau_{[\mu}{}^{+} e_{\nu]}{}^{a'} (S^{(-)}_{+}{}^{b' c'} \partial_{a'}\Lambda_{b' c'} - S^{(-)}_{a' {b' c'}} \partial_{+}\Lambda^{d' c'}) - e_{[\mu}{}^{a'} e_{\nu]}{}^{b'} S^{(-)}{}_{a'}{}^{d' c'} \partial_{b'} \Lambda_{d' c'} \nn \\ && - 2 (\tau_{[\mu}{}^- \tau_{\nu]}{}^{+} S^{(-)}_{+b'c'} + \tau_{[\mu}{}^{-} e_{\nu]}{}^{a'} S^{(-)}_{a'b'c'} )\partial_{-} \Lambda^{b' c'} \, .
\label{gravGSBR}
\eea
Similarly, one can use the identification on $F_{\mu \nu}{}^{i}$ to obtain
\bea
F^{(-)}_{\mu \nu b' c'} & = & \sqrt{2} \partial_{[\mu}(\tau_{\nu]}{}^{+} S^{(-)}_{+b'c'} + e_{\nu]}{}^{a'} S^{(-)}_{a' b' c'}) \nn \\
&&
- \sqrt{2} (\tau_{[\mu}{}^{+} S^{(-)}_{+[b'|}{}^{d'} + e_{\mu}{}^{a'} S^{(-)}_{a' [b'|}{}^{d'}) (\tau_{\nu]}{}^{+} S^{(-)}_{+d'|c']} + e_{\nu}{}^{e'} S^{(-)}_{e' d'|c']}) \, .
\eea
Part of the finite contributions to the gravitational action in the Bergshoeff-Romano formulation are given by the square of the previous object, while other contributions should be constructed by considering the identification over the full finite action (see appendix B of \cite{LescanoHet}). The main difference with the previous section is that in the case $v_{-}{}^{i}=0$, the gauge GS mechanism cannot be trivialized.
Therefore, relaxing the previous condition and finding the complete form of the gravitationally induced corrections is a promising way to further develop the identification in this formalism.

\section{Leading order boost transformations in the presence of gauge fields}
Our strategy to compute the higher-derivative boost corrections in a systematic way is as follows: first, we compute the gauge corrections to the boost transformations (this section) and then we transform these corrections by imposing the non-relativistic version of the BdRi (next section). There are two sources of gauge corrections: the one coming from the construction of the gauge contributions (Double Field Theory) and an extra one, due to the trivialization of the $SO(8)$ GS mechanism.

\subsection{Computation of the boost transformations from Gauged Double Field Theory}

The boost transformations for the gauge fields can be easily read from non-relativistic DFT. In order to have compatibility with the higher-derivative gravitational part, we need the following boost transformations
\bea
\delta_{\lambda} \alpha_{a i} & = &0 \, , \\
\delta_{\lambda} a_{\mu}^{i} & = & - \delta_{\lambda} S_{\mu}{}^{a' b'} (t^i)_{a' b'} \, ,
\eea
These transformations can be imposed by paying the cost of deforming the boost transformations of both $h_{\mu \nu}$ and $b_{\mu \nu}$. At leading order, the generalized metric is \cite{EandD}
\begin{align}    
    {\mathcal{H}}^{(0)}_{\cal MN} &= \begin{pmatrix}
	 h^{\mu \nu} 	 &  - h^{\mu \rho} {c}_{\rho \nu} + y^{\mu}_a x_{\nu}^a & 0 \\ - h^{\nu \rho} {c}_{\rho \mu} + y^{\nu}_a x_{\mu}^a
		   &   h_{\mu\nu} + 2 a_{\mu\nu} - {c}_{\rho \mu} h^{\rho \sigma} {c}_{\sigma \nu} + 2  {c}_{\rho (\mu} x_{\mu)}^a y^{\rho}_a & 0 \\ 0 & 0 & \kappa^{ij} 
	\end{pmatrix}, \label{eq:HeteroticGMLeadingOrder}
\end{align}
with $x_\mu^a = {\tau_\mu}^a$ and $y^\mu_c = - {\tau_a}^\mu \eta^{ab} (\alpha_{bc} - \epsilon_{bc})$ and $\alpha_{a b} = \frac12 \alpha_{a}{}^{i} \alpha_{b i}$. Every component of the generalized metric is boost invariant. So we can explicitly find the corrections to $b_{\mu \nu}$ under gauged boost transformations,  
\bea
\delta_{\lambda} b_{\mu \nu} = - 2 \epsilon_{a b} \lambda^{a}{}_{a'} \tau_{[\mu}{}^{b} e_{\nu]}{}^{a'} + \Delta_{\lambda} b_{\mu \nu} \, .
\eea
By forcing $\delta_{\lambda} {\cal H}_{\mu}{}^{\nu}=0$ we find
\bea
\Delta_{\lambda} b_{\rho \nu} = \tau_{[\nu}{}^{a} \delta S_{\rho] a' b'} (t_{i})^{a' b'} \alpha_{a}^{i} + e_{[\rho}{}^{a'} \lambda_{c a'} \eta^{c b} \alpha_{b a} \tau_{\nu]}{}^{a} \, . 
\label{gaugedboostb}
\eea
Repeating the procedure for $\delta_{\lambda} {\cal H}_{\mu \nu}=0$ we get
\bea
\delta_{\lambda} h_{\mu \nu} & = & - 2 \tau_{(\mu}{}^{a} \delta_{\lambda} a_{\nu)}{}^{i} \alpha_{a i} + 2 \delta_{\lambda} b_{\rho (\mu|} h^{\rho \sigma} b_{\sigma |\nu)} + \delta_{\lambda} b_{\rho (\mu} h^{\rho \sigma} \tau_{\nu)}{}^{a} a_{\sigma}{}^{i} \alpha_{a i} 
\nn \\ &&
+ b_{\rho (\mu} h^{\rho \sigma} \tau_{\nu)}{}^{a} \delta_{\lambda} a_{\sigma}{}^{i} \alpha_{a i} + \frac12 \tau_{\mu}{}^{a} \delta_{\lambda} a_{\rho}{}^{i} \alpha_{ai} h^{\rho \sigma} \tau_{\nu}{}^{b} a_{\sigma}{}^{j} \alpha_{b j} 
\\ \nn &&
+ [\delta_{\lambda} b_{\rho (\mu} \tau_{\nu)}{}^{a} \tau^{\rho}{}_{b} + b_{\rho (\mu} \tau_{\nu)}{}^{a} \delta_{\lambda}\tau^{\rho}{}_{b} + \frac12 \delta_{\lambda} a_{(\mu}{}^{i} \alpha_{b i} \tau_{\nu)}{}^{a} 
\nn \\ && + \frac12 \tau_{(\mu}{}^{d} \delta_{\lambda} a_{\rho}{}^{i} \alpha_{d i} \tau_{\nu)}{}^{a} \tau^{\rho}{}_{b} + \frac12 \tau_{(\mu}{}^{d} a_{\rho}{}^{i} \alpha_{d i} \tau_{\nu)}{}^{a} \delta_{\lambda} \tau^{\rho}{}_{b}] \eta^{b c} (\alpha_{c a} - \epsilon_{c a}) \, , \\
& = & - 2 \lambda_{a a'} e_{(\mu}{}^{a'} \tau_{\nu)}{}^{a} + \Delta_{\lambda} h_{\mu \nu}
\eea
where  $\delta_{\lambda} \tau^{\mu}{}_{a}=e^{\mu}{}_{a'} \lambda_{a}{}^{a'} + \Delta_{\lambda} \tau^{\mu}{}_{a}$. 

In order to find the final form of the $\delta_{\lambda} \tau^{\mu}{}_{a}$ transformation, we recall the constraint
\bea
(\Delta_{\lambda} h_{\mu \nu}) \tau^{\nu}{}_{a} + h_{\mu \nu} (\Delta_{\lambda} \tau^{\nu}{}_{a}) = 0 \, . 
\eea
Multiplying by $h^{\mu \rho}$ we get,
\bea
(\Delta_{\lambda} h_{\mu \nu}) \tau^{\nu}{}_{a} h^{\mu \rho} +\Delta_{\lambda} \tau^{\nu}{}_{a} (\delta_{\nu}^{\rho} - \tau_{\nu}{}^{b} \tau^{\rho}{}_{b}) = 0 \, , 
\eea
and therefore, all the new gauge contributions in $\delta_{\lambda} \tau^{\rho}{}_{a}$ are given by 
\bea
\Delta_{\lambda} \tau^{\rho}{}_{a} = - \Delta_{\lambda} h_{\mu \nu} \tau^{\nu}{}_{a} h^{\mu \rho} \, . 
\eea
\subsection{Induced boost Green-Schwarz mechanism coming from the trivialization of the gauge mechanism}
The gauge transformation of the b-field is given by, 
\bea
 \delta_{ \lambda} b_{\mu\nu} & = & -(\partial_{[\mu} \lambda^i) \alpha_{-i} \tau_{\nu]}^- \label{gaugetransb} \, ,
\eea
where $\alpha_{-}^i$ transforms as a gauge vector. Therefore, $b_{\mu\nu}$ contains a GS mechanism. While in the relativistic case this transformation is non-ambiguous and cannot be removed, here it is possible to define the field 
\bea
\bar b_{\mu\nu} = b_{\mu\nu} + a_{[\mu| i} \alpha_{-}^i \tau_{|\nu] -}
\label{tildeb}
\eea
which is gauge invariant, i.e. $\delta_{\lambda}\bar b_{\mu \nu}=0$. This is an interesting property of the heterotic supergravity under the NR limit, but there is a price to pay. The trivialization of the gauge GS mechanism induces a new boost transformation,
\bea
\delta_{\lambda} \bar b_{\mu\nu} = \delta_{\lambda} b_{\mu\nu} + \delta_{\lambda} a_{[\mu| i} \alpha_{-}^i \tau_{|\nu] -} \, .
\label{extragaugeinduced}
\eea
This new contribution cannot be trivialized, so the anomaly cancellation principle can be constructed from a pure gauge transformation, or, equivalently, with a gauge boost transformation. We will discuss the implications of this in the discussion section.

\section{Higher-derivative boost transformations}
\label{Sec:newboosts}

In this section we will compute the hidden higher-derivative boost transformations of the heterotic supergravity.  In the subsection \ref{Extra1} we compute the four-derivative boost transformations of the gravitational sector of the four-derivative heterotic supergravity. In particular, these contributions do not depend on any trivialization and are a direct consequence of the BdRi. In the subsection \ref{Extra2} we explicitly show that the trivialization of the $SO(8)$ GS mechanism produces an induced boost GS mechanism that cannot be further trivialized.

\subsection{Hidden boost transformations}
\label{Extra1}

While the $SO(8)$ GS mechanism rules the higher-derivative correction of the b-field after considering the NR limit, there exist extra higher-derivative corrections coming from the boost transformations that have been omitted in \cite{Trivial}. At this point we will forget about the trivialization of the $SO(8)$ GS mechanism and, keeping this mechanism on, we will explore what the boost transformations of the gravitational fields are. For doing so, we will first explore the boost transformations of the gauge fields and then we will impose the NR BdRi.

The BdRi on $\Delta_{\lambda} b_{\mu \nu}$, $\Delta_{\lambda} h_{\mu \nu}$ and $\Delta_{\lambda} \tau^{\mu}{}_{a}$  produce higher-derivative boost transformations given by
\bea
\delta^{(1)}_{\lambda} b_{\rho \nu} = \tau_{[\nu}{}^{+} \delta^{(0)}_{\lambda} S_{\rho] a' b'} \beta_{+}{}^{a' b'} - e_{[\rho}{}^{a'} \lambda_{c a'} \eta^{c b} \beta_{+}^{2} \tau_{\nu]}{}^{a} \, , 
\label{higherboostb}
\eea
where
\bea
\delta_{\lambda} S_{\mu}{}^{a' b'} & = & 2 e^{\nu [a'} \partial_{[\mu} (\tau_{\nu]}{}^{b} \lambda_{b}{}^{b']}) - \tau_{\mu}{}^{b} \lambda_{b c'} e^{\nu a'} e^{\rho b'} \partial_{[\nu} e_{\rho]}{}^{c'} \nn \\ && 
- e_{\mu c'} e^{\nu a'} e^{\rho b'} \partial_{[\nu} \tau_{\rho]} \lambda_{b}{}^{c'} + 3 \tau_{\mu}{}^{a} \epsilon_{a b} \tau^{\nu b} e^{\rho a'} e^{\sigma b'} \partial_{[\nu} (\epsilon_{c d} \lambda^{c}{}_{c'} \tau_{\rho}{}^{d} e_{\sigma]}{}^{c'}) 
\nn \\ &&
- \frac12 \tau_{\mu}{}^{a} \epsilon_{a b} e^{\nu}{}_{c'} \lambda^{b c'} e^{\rho a'} e^{\sigma b'} h_{\nu \rho \sigma} \, .
\eea
The full boost transformation for $h_{\mu \nu}$ is given by
  \bea
\delta^{(1)}_{\lambda} h_{\mu \nu} & = & \frac{2}{\sqrt 2} \tau_{(\mu}{}^{a} \delta^{(0)}_{\lambda} S_{\nu)}{}^{c' d'} \beta_{a c' d'} + 2 \delta^{(1)}_{\lambda} b_{\rho (\mu|} h^{\rho \sigma} b_{\sigma |\nu)} - \frac{1}{\sqrt{2}} \delta^{(0)}_{\lambda} b_{\rho (\mu} h^{\rho \sigma} \tau_{\nu)}{}^{a} S_{\sigma}{}^{c' d'} \beta_{a c' d'} 
\nn \\ &&
- \frac{1}{\sqrt 2} b_{\rho (\mu} h^{\rho \sigma} \tau_{\nu)}{}^{a} \delta^{(0)}_{\lambda} S_{\sigma}{}^{c' d'} \beta_{a c' d'}
- [\delta^{(1)}_{\lambda} b_{\rho (\mu} \tau_{\nu)}{}^{a} \tau^{\rho}{}_{b} + b_{\rho (\mu} \tau_{\nu)}{}^{a} e^{\rho}_{a'} \lambda_{b}{}^{a'} 
\\ \nn &&
- \frac{1}{2 \sqrt2} \delta^{(0)}_{\lambda} S_{(\mu}{}^{c' d'} \beta_{b c' d'} \tau_{\nu)}{}^{a} 
- \frac{1}{2\sqrt 2} \tau_{(\mu}{}^{d} \delta^{(0)}_{\lambda} S_{\rho}{}^{c' d'} \beta_{d c' d'} \tau_{\nu)}{}^{a} \tau^{\rho}{}_{b} 
\\ \nn && 
- \frac12 \tau_{(\mu}{}^{d} S_{\rho}{}^{c' d'} \beta_{d c' d'} \tau_{\nu)}{}^{a} e^{\rho}{}_{a'} \lambda_{b}{}^{a'}] \eta^{b c} (\beta_{c a} + \epsilon_{c a}) \, .
\label{higherboosth}
\eea
From the previous expression, it is straightforward to obtain $\delta^{(1)}_{\lambda} \tau^{\nu}{}_{a}$ by using
\bea
\delta^{(1)}_{\lambda} \tau^{\rho}{}_{a} = - \delta^{(1)}_{\lambda} h_{\mu \nu} \tau^{\nu}{}_{a} h^{\mu \rho} \, . 
\eea.

As was mentioned in \cite{Trivial}, the four-derivative gravitational action is not boost-invariant, since the leading order transformations acting on this action compensate the action of the hidden transformations on the leading order action.

\subsection{Induced boost Green-Schwarz mechanism coming from the trivialization of the $SO(8)$ mechanism}
\label{Extra2}
Let's continue by exploring the transformation of the new field-redefined field, $\bar b_{\mu \nu}$ under boost transformations,

\bea
\delta_{\lambda} \bar b_{\mu \nu} = \delta_{\lambda} b_{\mu \nu} +  \frac{5}{12} \delta_{\lambda}S^{(-)}_{[\mu}{}^{a' b'} \tau_{\nu]}{}^{+}  \tau_{\rho \sigma}{}^{-}  e^{\rho}{}_{a'} e^{\sigma}{}_{b'} \, .
\label{extragravityinduced}
\eea

The previous contribution constitutes a boost-induced GS mechanism, which further extends the transformation of the previous section,
\bea
\delta_{\lambda} \bar b_{\mu \nu} = - 2 \epsilon_{a b} \lambda^{a}{}_{a'} \tau_{[\mu}{}^{b} e_{\nu]}{}^{a'} +  \Delta_{\lambda} b_{\mu \nu} +  \frac{5}{12} \delta_{\lambda}S^{(-)}_{[\mu}{}^{a' b'} \tau_{\nu]}{}^{+}  \tau_{\rho \sigma}{}^{-}  e^{\rho}{}_{a'} e^{\sigma}{}_{b'}
\eea
The next question is whether this new mechanism might be trivialized without including further transformations and the answer is no. While some of the contributions can be eliminated by a field redefinition of the boost parameter, 
\bea
\delta^{(1)}_{\tilde \lambda} b_{\mu \nu} = - 2 \epsilon_{a b} \tilde \lambda^{a}{}_{a'} \tau_{[\mu}{}^{b} e_{\nu]}{}^{a'} \, ,
\eea
with $\tilde \lambda$ a field redefinition, in general this transformation is a genuine non-trivial boost-induced GS mechanism. 

The mechanism shown in this section has appeared in order to compensate the trivialization of the $SO(8)$ GS mechanism. However, further higher-derivative boost transformations should be already present even without imposing the trivialization

\section{Discussion: Towards a Metsaev-Tseytlin formulation of NR heterotic string}

The four-derivative corrections to the universal action principle for the different formulations of closed string theory were historically computed considering three- and four-point scattering amplitudes for the massless states \cite{GrossSloan} \cite{CN}. This method is based on the study of the different string interactions through the S-matrix, in order to construct an effective Lagrangian which would be able to reproduce those interactions. The effective action, originally computed by Metsaev and Tseytlin, takes the form
\bea
S_{MT} = \int d^Dx \sqrt{-g} e^{-2\phi} (L^{(0)} + L^{(1)}_{MT}) \, ,
\label{fullA}
\eea
where $L^{(0)}$ is the two-derivative NSNS supergravity Lagrangian and
\bea
L^{(1)}_{\rm MT} & = & \frac{1}{8} \Big[  R_{\mu \nu \rho \sigma} R^{\mu \nu \rho \sigma} - \frac12 H^{\mu \nu \rho} H_{\mu \sigma \lambda} R_{\nu \rho}{}^{\sigma \lambda} + \frac{1}{24} H^4 - \frac18 H^2_{\mu \nu} H^{2 \mu \nu} \Big] \nn \\ && - \frac{1}{4} H^{\mu \nu \rho} C_{\mu \nu \rho} \, .
\label{MT}
\eea
In the previous expression we have introduced the following notation,
\bea
H^2_{\mu \nu} & = & H_{\mu}{}^{\rho \sigma} H_{\nu \rho \sigma} \, , \\
H^2 & = & H_{\mu \nu \rho} H^{\mu \nu \rho} \, .
\eea
The invariance under Lorentz transformations of the MT action requires a modification of the Lorentz transformation of the $b$-field given by,
\bea
\delta_{\Lambda} B_{\mu \nu} = -\frac12 \partial_{[\mu} \Lambda^{a b} w_{\nu] a b} \, .
\eea
In order to connect this action with the action of Bergshoeff and de Roo, one needs to impose the following non-covariant field redefinition on the B-field,
\bea
B^{\rm{MT}}_{\mu \nu} = B^{\rm{BdR}}_{\mu \nu} - \frac14 H_{[\mu}{}^{ab} w_{\nu]ab} \, .
\label{bequivalence}
\eea
While in the relativistic case both formulations are physically the same (see \cite{Tduality} for the full proof), the non-relativistic limit of the action (\ref{MT}) has not yet been obtained. The Riemann squared contribution contains divergences \cite{CR} that should be canceled by the Chern-Simons contributions as happens in the formulation constructed here using the BdRi. For example, our Lagrangian does not contain terms given by $R^{(-)}_{\mu \nu a' b'} R^{(-) \mu \nu a' b'}$ since in the two-derivative heterotic Lagrangian we do not have $F^2$ terms after considering the NR limit. This cancellation of divergences in the Metsaev and Tseytlin formulation in the NR limit could be related to field redefinitions equivalent to (\ref{bequivalence}) in this limit or to higher-derivative extensions of the non-relativistic expansion of the fundamental fields. Finally, the construction of the Metsaev and Tseytlin action could allow us to obtain the form of the gravitational four-derivative corrections for the bosonic string, so it constitutes a relevant immediate direction of research.

\section{Conclusions and outlook}

In this work, we have constructed the four-derivative extension of NR heterotic supergravity by developing a consistent NR generalization of the BdRi. We showed that the non-Abelian gauge transformations of the heterotic theory match the structure of the gravitational GS anomaly cancellation mechanism when a suitable light-cone decomposition of the Newton-Cartan geometry is employed. This matching required a specific field redefinition of the Kalb-Ramond field and led to the identification of the gauge field as an $SO(8)$ spin connection with torsion. With this identification, the three-form field strength acquires a Chern-Simons-like correction compatible with the NR symmetry structure. Using this framework, we derived the finite four-derivative corrections to the NR heterotic action. The resulting action is manifestly invariant under $SO(8)$ transformations, $SO(1,1)$  transformations and diffeomorphisms. While the leading two-derivative NSNS action enjoys boost invariance, the four-derivative terms generated via the Bergshoeff-de Roo procedure require higher-derivative corrections to these transformations. 

This new set of higher-derivative boost transformations provide a useful perspective on the Green--Schwarz mechanism in the non-relativistic limit. In relativistic heterotic supergravity, anomaly cancellation is implemented through the transformation properties of the two-form under gauge and local Lorentz transformations. The non-relativistic limit changes the realization of these symmetries in an essential way. In particular, transformations that are non-trivial in the relativistic theory can become trivializable after taking the limit and making suitable field redefinitions. It is important, however, that such a trivialization should not be interpreted as the disappearance of the Green--Schwarz mechanism. Rather, the anomaly-cancellation structure is redistributed among the symmetries of the non-relativistic theory.

While the gravitational case is independent of the gauge one, the exact mechanism appears in the formulation constructed in \cite{LescanoHet}. This gauge Green--Schwarz transformation can be removed by introducing a gauge-invariant redefined two-form. In this set of variables, the two-form no longer transforms under the non-Abelian gauge symmetry. The price for this simplification is that the boost transformation of the redefined field acquires an additional contribution involving the gauge connection as shown in (\ref{extragaugeinduced}). Thus, the field redefinition does not remove the Green--Schwarz structure, but transfers it from the gauge symmetry to the boost symmetry, as shown in (\ref{extragravityinduced}). 

A similar mechanism occurs in the gravitational sector. The non-relativistic $SO(8)$ Green--Schwarz transformation can be partially trivialized by a field redefinition involving the torsionful connection \cite{Trivial}. This redefinition is particularly useful because it allows the gravitational contribution to be related directly to the gauge contribution through the non-relativistic Bergshoeff--de Roo identification. At the same time, the redefinition induces an additional transformation under boosts. Consequently, the gravitational Green--Schwarz mechanism is likewise not eliminated by the choice of variables. Instead, part of its realization is transferred to the boost symmetry.

These observations suggest that anomaly cancellation has to be understood somewhat differently in the non-relativistic theory. The non-relativistic limit does not remove the need for a Green--Schwarz mechanism. Rather, it changes the symmetry in which this mechanism is realized. Once the gauge and Lorentz Green--Schwarz transformations are trivialized, their information has to be carried by the remaining non-relativistic transformations. In the formulation considered here, this role is played by the boosts. The resulting boost Green--Schwarz transformations therefore provide an essential part of the non-relativistic realization of anomaly cancellation. There are consequently two contributions to this structure. The first originates from the gauge sector and is induced when the gauge Green--Schwarz transformation is trivialized. The second originates from the gravitational sector and appears when the $SO(8)$ Green--Schwarz transformation is reorganized through the torsionful connection and the corresponding field redefinition. The non-relativistic Bergshoeff--de Roo identification relates these two sectors and shows that they should not be regarded as independent modifications of the boost algebra. Instead, they represent different origins of the same underlying Green--Schwarz structure inherited from heterotic supergravity.

We therefore expect anomaly cancellation in non-relativistic heterotic supergravity to be realized through a generalized Green--Schwarz mechanism involving the boost symmetry. Both the gauge and gravitational sectors contribute to this mechanism. From this viewpoint, the higher-derivative boost transformations found here are an intrinsic part of the non-relativistic heterotic theory. A complete understanding of the anomaly-cancellation mechanism in the non-relativistic theory should consequently include the boost sector on the same footing as the gauge and local Lorentz sectors.

\subsection{Outlook}

The results of this work provide a finite formulation of higher-curvature dynamics in NR string backgrounds and open the door to further studies such as: 
\begin{itemize}
\item {\bf Non-relativistic four-derivative bosonic supergravity}: As discussed in the previous section, one important next step in this program is the computation of the NR limit of the action (\ref{MT}) and the proof, if it exists, an equivalence with the formulation of Bergshoeff-de Roo (see \cite{New} for recent advances in this direction). In the relativistic case (\ref{MT}) is related to the BdRi via non-covariant field redefinitions \cite{Review}. It is not clear a priori if these two formulations can be related, after taking the NR limit, using field redefinitions.   
\item {\bf Inclusion of ${\cal N}=1$ Supersymmetry}: Another possibility is to incorporate the fermionic sector of the theory, following the lines of the original formulation of Bergshoeff and de Roo \cite{BdR}, in the NR limit. Initial steps in this direction were given in \cite{Bsusy} by considering the leading order in fermions. While the supersymmetry transformations diverge in this prescription, starting from $N=1$ DFT could bring new possibilities to explore these transformations using a more general set of fields.

\item {\bf T-duality invariant rewriting}: Finally, one interesting possibility is to uplift the present formulation to the framework of DFT \cite{DFT1}-\cite{DFT2}, in a non-Riemannian formulation \cite{NRDFT1}-\cite{NRDFT5}, with higher-derivative terms following the construction given in \cite{HSZ}. It would be very interesting to incorporate the technology of the generalized BdRi constructed in \cite{gBdR}-\cite{LMG} \footnote{See also \cite{Achilleas} for an alternative approach to this topic.} in the NR limit of DFT given in \cite{EandD}. 
\end{itemize}  

\subsection*{Acknowledgements}
The author is very grateful to D. Marques and J. Rosseel for discussions, and to D. Osten for partial collaboration in the computation of the higher-derivative structure of the boost transformations.  E.L. is supported by the Slovenian Quantum Science Hub — SQUASH programme, cofunded by the European Union under the Horizon Europe Marie Sklodowska-Curie Actions COFUND programme.

{}
\end{document}